\documentclass{elsart}
\usepackage{epsf}
\begin{document}
\begin{frontmatter}
\title{Bond-orientational ordering and shear rigidity in modulated 
colloidal liquids}
\author{Chinmay Das, A. K. Sood and H. R. Krishnamurthy}
\address{Department of Physics, Indian Institute of Science,
Bangalore, India, 560 012}
\begin{abstract}
From Landau-Alexander-McTague theory and Monte-Carlo simulation results
we show that the {\em modulated liquid} 
obtained by subjecting a colloidal system to a periodic
laser modulation  
has long range bond-orientational order 
and non-zero shear rigidity. 
From {\em infinite field} simulation results we show that in the 
{\em modulated liquid} phase, the translational order parameter 
correlation function decays to zero exponentially while the 
correlation function for the bond-orientational order saturates 
to a finite value at large distances.
\end{abstract}
\end{frontmatter}

\section{Introduction}
Consider a 2-d charge stabilized colloidal system subject to  
a one-dimensionally modulated stationary laser field (obtained by
superposing  two beams). 
Chowdhury, Ackerson and Clark \cite{expt:ackerson}
 found that when the wave-vector of the laser modulation
is tuned to be half the wave-vector $q_0$ at which
the structure factor of the colloidal liquid shows its
first peak,
above a certain laser field intensity,
the system freezes into  a 2-D triangular crystal. 
This phenomena
of laser induced freezing (LIF)
has been studied subsequently by experiments [2-4], simulations[5-7] 
and density functional
calculations [8-10].
For field strengths below the value at which
the system undergoes freezing, the laser field induces a
density modulation with wave-vector $q_0$ in the liquid, and this phase
has been
called
{\it modulated liquid} in the literature. 
In this paper we show that the modulated liquid has rather interesting
properties. Specifically
the induced translational order generates a field conjugate to the
bond-orientational order parameter. Hence the {\it "modulated liquid"}
has a non-zero value of bond-orientational order parameter, and consequently
a finite rigidity (shear modulus).
We first present a qualitative picture in terms of
a generalized Landau-Alexander-McTague \cite{th:landau} 
theory and substantiate
the picture with results from a detailed Monte-Carlo simulation.

\section{Mean-field treatment}
From general symmetry grounds the coarse-grained
free energy functional in two
dimensions in the presence of an external field $V_e$, coupled to 
one of the six density Fourier modes $(\rho_{\vec{G_1}})$ characterizing
a 2-D crystal has the 
form \cite{nelson}:
\begin{eqnarray}
F &=& - 2 V_e \rho_{\vec{G_1}} + 
\frac{1}{2} r_T \sum_{\vec{G}} |\rho_{\vec{G}}|^2
\nonumber \\
&& 
+ w_T \sum_{\vec{G} + \vec{G'} + \vec{G''} = 0}
\rho_{\vec{G}} \rho_{\vec{G'}} \rho_{\vec{G''}}
+ u_T (\sum_{\vec{G}} |\rho_{\vec{G}}|^2)^2
+ u_T' \sum_{\vec{G}} |\rho_{\vec{G}}|^4
+ ... \nonumber\\
& & + \frac{1}{2} r_6 |\psi_6|^2 + u_6 |\psi_6|^4 + ... \nonumber\\
& & + \gamma \sum_{\vec{G}} |\rho_{\vec{G}}|^2 
[\psi_6 (G_x - i G_y)^6 + \psi_6^* (G_x + i G_y)^6] \label{eq:landau};
\end{eqnarray}
where the summations run over all the wave-vectors forming the first shell
of the reciprocal lattice vector of the triangular lattice into which the
colloid freezes
and $\psi_6$ is the bond-orientational order parameter.

\begin{figure}[h]
\parbox{0.4\linewidth}{
\vbox{\epsfxsize=3cm
\epsfbox{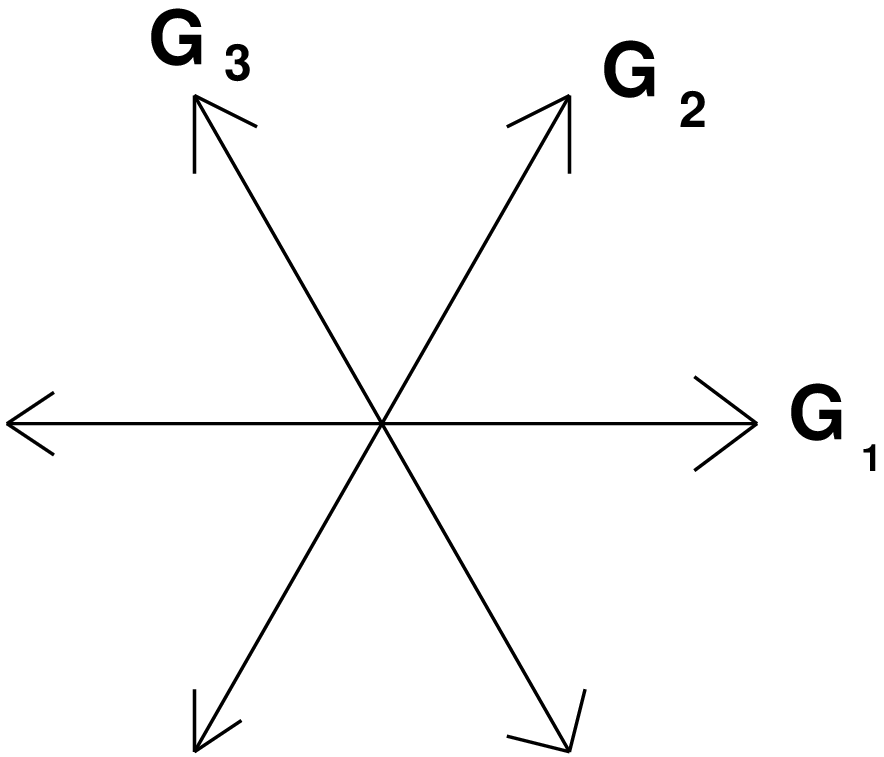}
}}
\hfill
\parbox{0.4\linewidth}{{\it Notations:}\\
$\rho_1 \equiv \rho_{\vec{G_1}} \equiv \rho_{-\vec{G_1}}$\\
$\rho_2 \equiv \rho_{\vec{G_2}} \equiv \rho_{\vec{G_3}}
\equiv \rho_{-\vec{G_2}} \equiv \rho_{-\vec{G_3}}$
}
\caption{Density Fourier modes for 2-D triangular crystal.}
\end{figure}

In the absence of external field, for large positive values of
$r_T$ and $r_6$, the free energy is minimum when all order parameters
are zero (ie. for the liquid phase). 
If, as a consequence of a change in temperature, screening length or some
other parameter, $r_T$ decreases much faster than $r_6$, then
because of the cubic term in equation (\ref{eq:landau}) one gets
a first order freezing transition into a crystalline phase.
On the other hand, if $r_6$ decreases much faster than $r_T$,
one gets a continuous transition to an orientationally
ordered hexatic phase,
characterized by nonzero values of $\psi_6$ and zero values of
$\rho_{\vec{G}}$. 

The external field induces a non-zero $\rho_{\vec{G_1}}$ even in
the liquid phase. The non-zero value of $\rho_{\vec{G_1}}$ leads to
an effective field conjugate to  the bond-orientational order through the
coupling $\gamma$ in eqn. \ref{eq:landau}. So $\psi_6$ is also turned on
as the external field is applied. Since the external field modulation
fixes the directions in which the order develops, we can treat the
order parameters as real number to get the equilibrium phase diagram
(the phase in the order parameters
 will be important in determining the elastic coefficients
but not in determining the stable equilibrium phase). The 
experimental systems and the earlier simulations 
referred to above [1-7] corresponded to 
volume fraction and salt concentration values such that the zero field
freezing transition was first order; 
hence they presumably correspond to
$r_T \leq r_6$.
Even if $r_6$ is much smaller than $r_T$, such that the 
freezing mechanism at zero external field
is a two stage transition with the first (continuous)
transition being to the hexatic phase, the
external field will immediately destroy the intervening hexatic phase.

\begin{figure}[htbp]
\epsfxsize=8cm
\centerline{\epsfbox{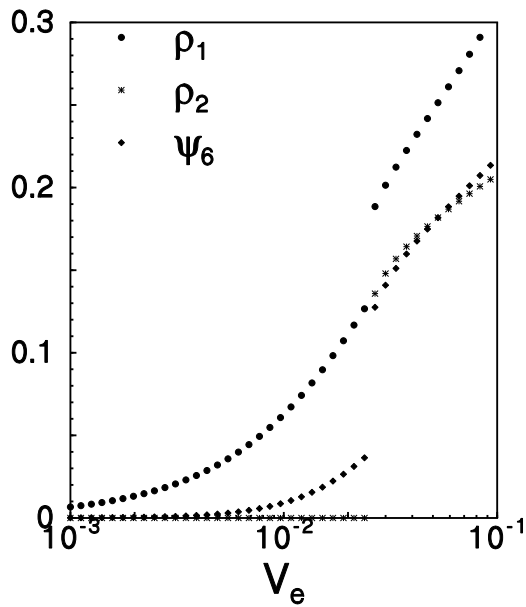}}
\caption{Values of the translational and 
orientational order parameters as a function
of external field in Landau-Alexander-McTague theory}
\label{fig:landau}
\end{figure}

Choosing the coefficients $r_T = r_6 = 0.30$, 
$w_T = - 1/3$, $u_T = u_6 = 1.5$, $u_T' = 0.5$ and 
$\gamma = -0.7$ 
numerical minimization of eqn. \ref{eq:landau} gives the order
parameter profile as a function of external field shown
in Fig.\ \ref{fig:landau}.
$\psi_6$ and $\rho_{\vec{G_1}}$ becomes nonzero as soon as
$V_e$ is turned on, and at
$V_e = 0.0239$ all the order parameters jump simultaneously signifying
a first order transition to a crystalline structure.

One can  define a rigidity modulus in analogy to the helicity modulus 
in superfluids \cite{fisher}:
\begin{equation}
Y = \lim_{q \rightarrow 0}\frac{1}{q^2}\frac{\partial^2 F}{\partial^2 \vec{q}},
\end{equation}
where we consider an field coupled to the orientational order
parameter as $\psi_6(\vec{r}) \longrightarrow \psi_6(\vec{r}) 
e^{i \vec{q} . \vec{r}}$. The free-energy cost for such a field
will depend on the gradient term in $\psi_6$ (not included
in eqn.\ \ref{eq:landau}).
To leading order  such a term would predict $Y \sim V_e^4$ as
$|q| \rightarrow 0$. 
\section{Monte-Carlo Simulations}
\subsection{Simulation details}
Since the parameters in eqn.\ \ref{eq:landau} are phenomenological  and there
is no direct way to fix them to characterize the experimental system,
we have performed Monte-Carlo simulations to study the {\it modulated
liquid phase}. We have considered a 2-D system of 
charge stabilized colloidal particles (with diameter
$2 R = 1.07 \mu m$) confined in a  rectangular box of size 
$ \frac{\sqrt{3}}{2} a_s L \times a_s L $ with periodic boundary conditions
and subjected to an external potential of the form
$U(\vec{r}) = - {V}_e cos(q_0 x)$, with $q_0 = 2 \pi /
(\frac{\sqrt{3}}{2} a_s)$,
where $a_s$ is the mean inter-particle separation.
The inter-particle
interaction is modeled by the DLVO potential:
\begin{equation}
U_{ij}(r) = \frac{(Z e)^2}{\epsilon}
(\frac{exp(\kappa R)}{1 + \kappa R})^2
\frac{exp(-\kappa r_{ij})}{|r_{ij}|}
\label{eq:DLVO}
\end{equation}
Here $Z e$ ($Z = 7800$) is the effective surface charge,
$\epsilon$ (=78) is the dielectric constant of the solvent and
$\kappa$ is the inverse of the Debye screening length due to
the small ions (counterions and impurity ions) in the solvent.
The parameters are the same as in earlier simulation studies and similar
to the experiments \cite{expt:ackerson}. 

To study the effect of very large external field ({\it infinite field}),
in some of the simulations we had fixed the particles in parallel lines
defined by the potential minima. Though the particles
 move freely only along the lines, they 
interact in full two dimensional space. The resulting simplification
allows us to simulate systems with $L$ as large as $100$, which allows
us to
compute correlation functions at large distances.

The translational order parameters
are defined by:
\begin{equation}
\rho_{\vec{G}} = <\frac{1}{N} \sum_i e^{i \vec{G} . \vec{r_i}}>. \label{eq:op}
\end{equation}
 $G_1$ refers to the
direction parallel to the modulation wave-vector $[\frac{2 \pi}{\sqrt{3}/2 a_s}
(1,0)]$, while $G_2$ refers to 
the other independent wave-vector 
$[\frac{2 \pi}{\sqrt{3}/2 a_s}
(\frac{1}{2}, \frac{\sqrt{3}}{2})]$ 
forming the first shell of wave-vectors
of the triangular lattice. 
Value of $\rho_{\vec{G}}$ as
defined in eqn.(\ref{eq:op}) depends on the coordinate origin. So we 
measure the translational order parameter as 
$\rho_{\vec{G}} = <\frac{1}{N} \sqrt{[\sum_i cos(\vec{G}.\vec{r_i})]^2
+ [\sum_i sin(\vec{G}.\vec{r_i})]^2}>$ \cite{numrec}.  This quantity
is of order unity in crystalline state, while in liquid this goes to
zero as $\frac{1}{\sqrt{N}}$ for a system of $N$ particles. 

We define the bond-orientational order parameter as:
\begin{equation}
\Psi_6 = <\frac{1}{N} \sum_m \frac{1}{z_m} \sum_n e^{6 i \theta_{m,n}}>,
\end{equation}
where $\theta_{m,n}$ is the angle made by the line joining the 
position of particle $m$ to the neighboring particle $n$, measured
with respect to some fixed direction. $z_m$ is the number of neighbors
corresponding to the Voronoi cell of particle $m$.

To calculate order parameter correlation functions we define
local order parameters $\rho_{\vec{G}} (\vec{r_m}) = 
\frac{1}{z_m+1} [e^{i \vec{G} . \vec{r_m}} + \sum_n e^{i \vec{G} . \vec{r_n}}]$
and $\psi_6(\vec{r_m}) = \frac{1}{z_m} \sum_n e^{6 i \theta_{m,n}}$, where
the order parameters are defined at the position of a particle $m$, and 
$n$ denotes the particles forming the Voronoi cell of particle $m$. The 
order parameter correlation functions are defined as,
\begin{eqnarray}
G_T(\vec{r}) & = &  <\rho_{\vec{G}}^*(\vec{r}) \rho_{\vec{G}}(0) >,\\
G_6(\vec{r}) &  = &  <\psi_6^*(\vec{r}) \psi_6(\vec{0})>.
\end{eqnarray}

To measure the helicity modulus $Y$, we have applied "anti-periodic" 
boundary condition along the $x$ direction. In case of the standard
periodic boundary condition,
one repeats the simulation cell throughout space for calculating
the inter-particle potential and to
put back the particle inside the simulation box once it moves out in the course
of the simulation.  In case of the "anti-periodic" boundary condition, 
in the x direction for example,
successive imaginary
repeat boxes 
are shifted by half the lattice spacing in the 
y-direction. 
Equivalently, while  folding back a particle that exits from
the simulation box in the $\pm x$ direction, a displacement
$\pm \frac{1}{2} a_s \hat{y}$ is applied. The change in energy
between periodic and "anti-periodic" boundary conditions gives
a measure of $Y$. 

\subsection{results}

\begin{figure}[htbp]
\epsfxsize=10cm
\epsfysize=6.2cm
\centerline{\epsfbox{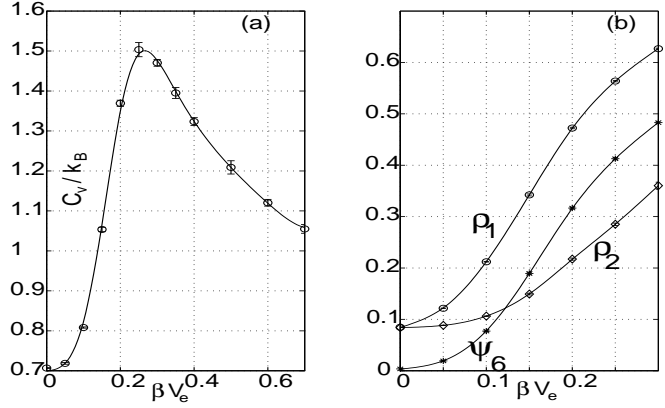}}
\caption{ (a) Specific heat, and (b) Order parameters, 
obtained from Monte-Carlo simulations of a 400 particle
system as a function of $\beta V_e$. $\kappa a_s$ was chosen to be
15.5 for these simulations, such that the system is in liquid state
in absence of external field}
\label{fig:simop}
\end{figure}

The specific heat in figure \ref{fig:simop} (a) shows a peak as function of
$\beta V_e$ signifying a phase transition at $\beta V_e = 0.25$.
In figure \ref{fig:simop}(b) we present order parameters  for
$L = 20$ and $\kappa a_s = 15.5$ (where the system is
liquid for zero external field) as a function of
$\beta V_e$. The translational order parameters $\rho_1$ and $\rho_2$
are non-zero even in the liquid phase because of the finite system size.
Also they seem to grow continuously. But a finite-size scaling analysis
shows that the results are consistent with first order transition scenario
with small discontinuity
 in energy \cite{yet_to_be}. 
In addition to $\rho_1$, $\Psi_6$ also
shows non-zero value as soon as the external field is switched on as
expected from the coarse-grained free energy.

\begin{figure}[htbp]
\centerline{\epsfxsize=5cm 
\epsfysize=5.9cm 
\epsfbox{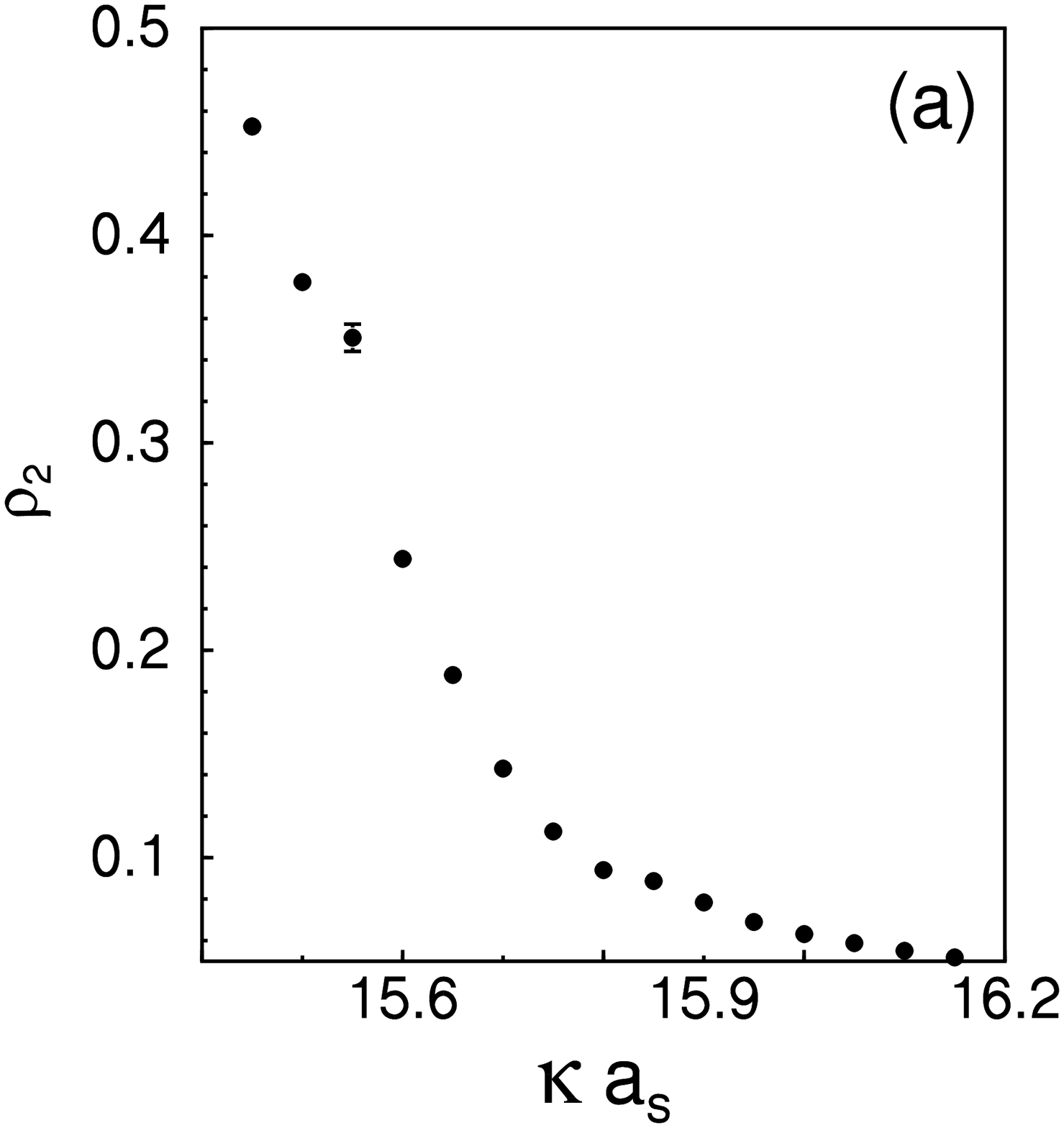} 
\epsfxsize=5cm
\epsfysize=5.9cm 
\epsfbox{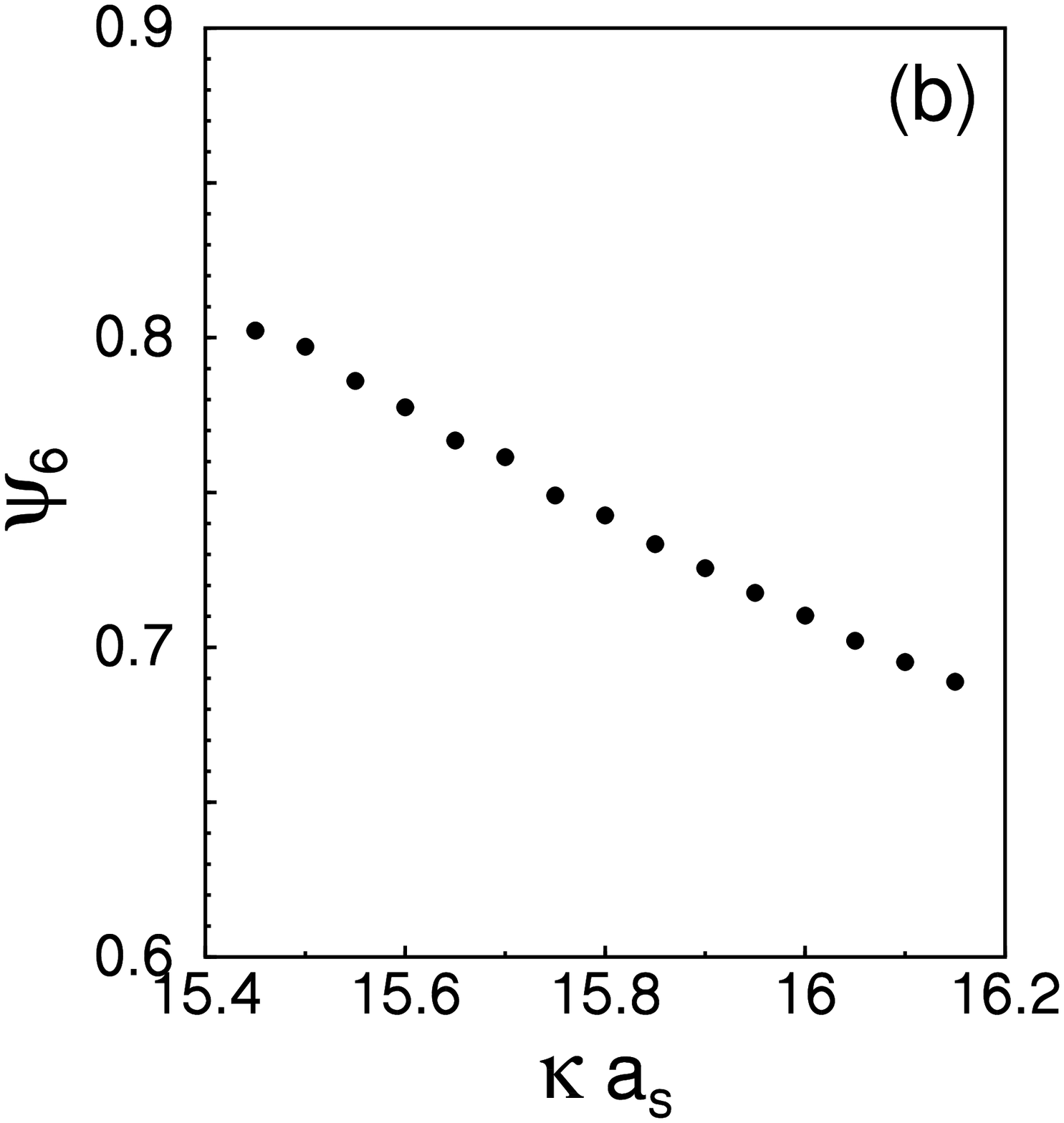}}
\caption{$\rho_{\vec{G_2}}$ and $\Psi_6$ across the melting transition
for infinite field and 10000 particles}
\label{fig:infop}
\end{figure}

In figure \ref{fig:infop}, 
we plot $\rho_2$ and $\Psi_6$ for infinite field
and $10^4$ particle system ($L = 100$) as a function of $\kappa a_s$. While
the translational order parameter shows a sharp fall at the melting
transition ($\kappa a_s = 15.6$), the 
bond-orientational order parameter remains large and finite.

\begin{figure}[htbp]
\centerline{
\epsfxsize=5cm
\epsfysize=6cm
\epsfbox{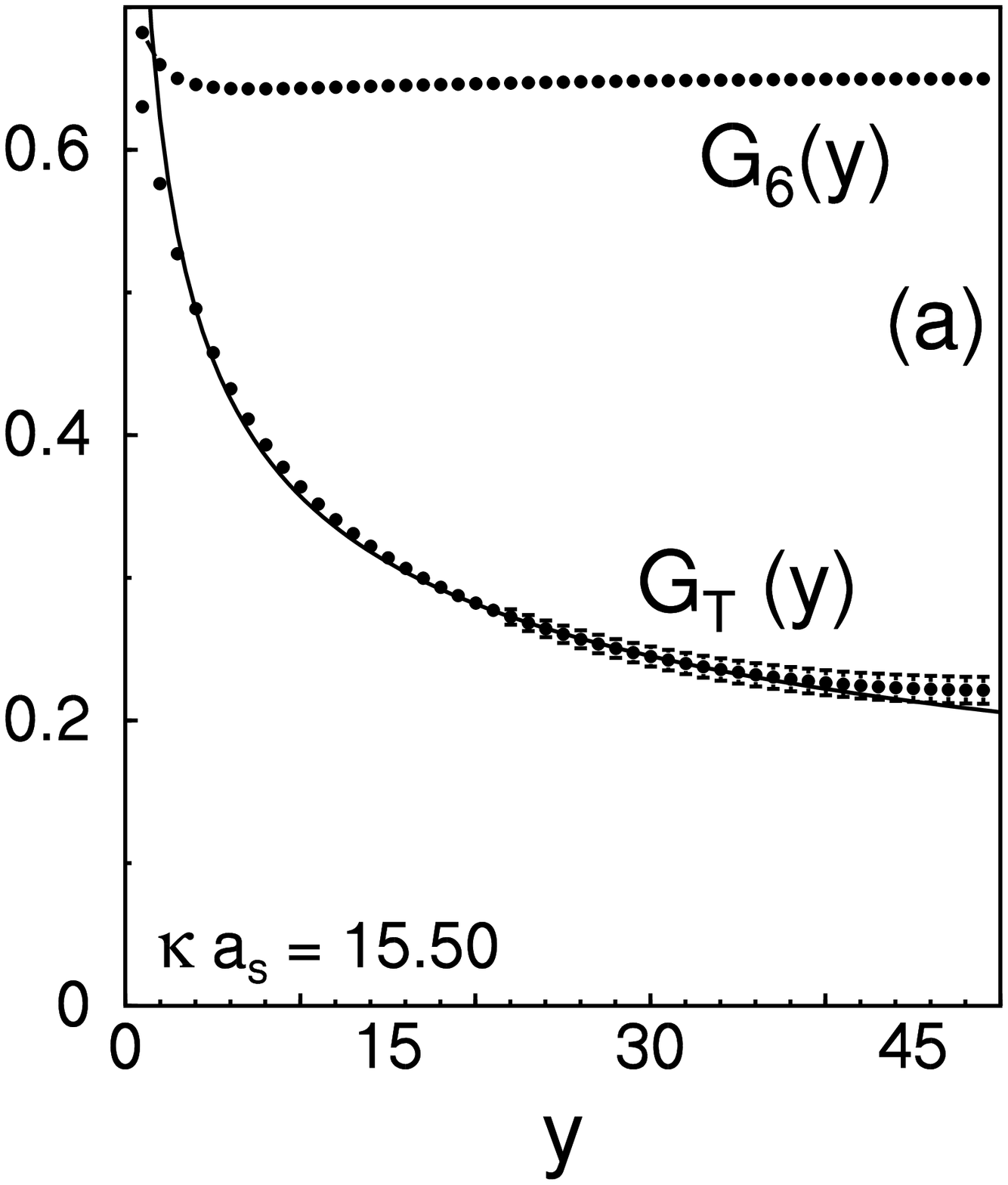}
\epsfxsize=5cm
\epsfysize=6cm
\epsfbox{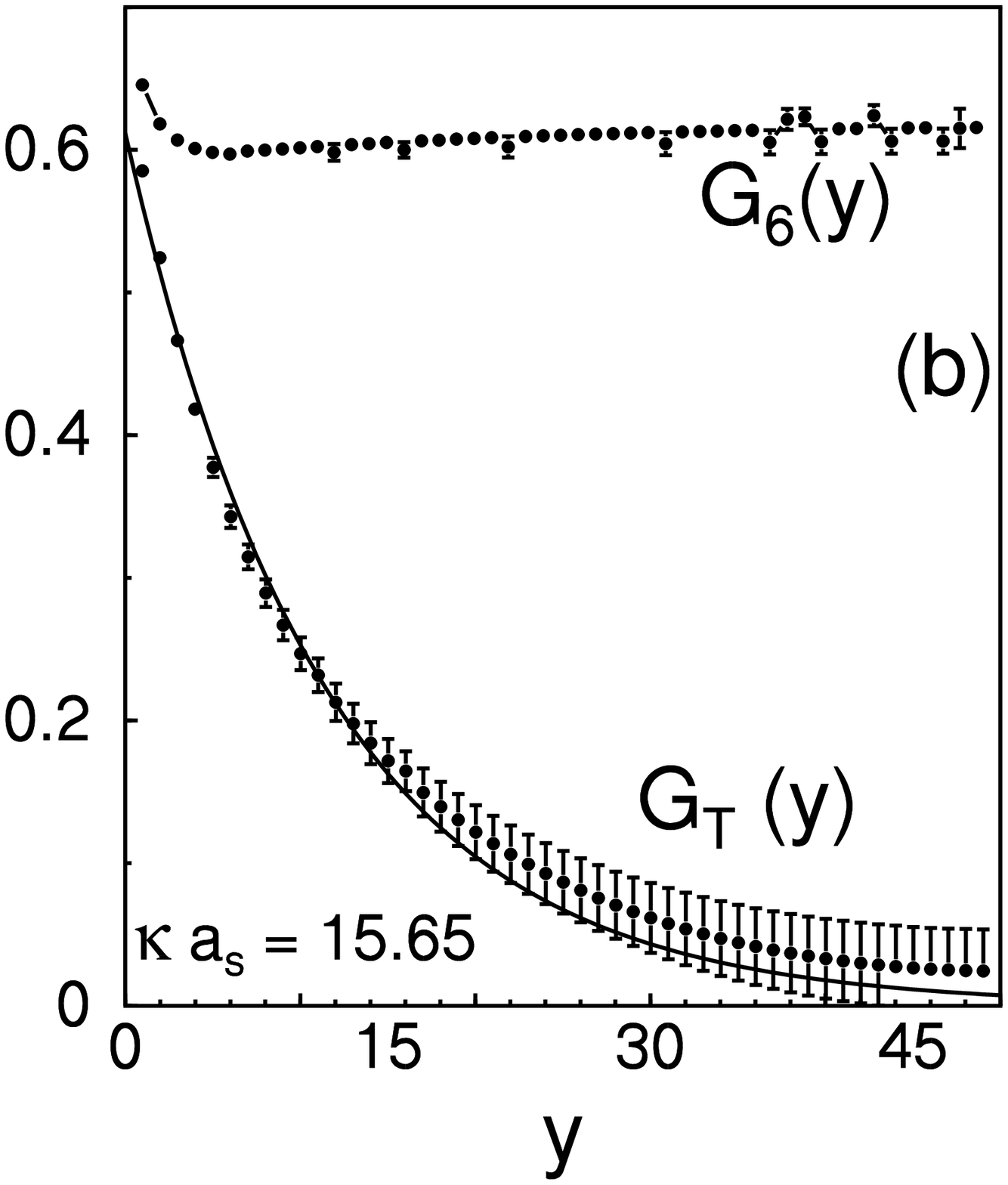}}
\caption{Translational and bond-orientational correlation functions
along the external field minima (y axis) from simulation of
10000 particles and infinite field}
\label{fig:simcor}
\end{figure}

In figure \ref{fig:simcor}
 we plot the order parameter correlation functions in the
crystalline region(\ref{fig:simcor}.a) and 
in the liquid region (\ref{fig:simcor}.b). In crystalline
region $G_T(y)$ decays as a power law, 
while $G_6(y)$ saturates
to a constant value. In the liquid region, while $G_T(y)$ decays to zero
exponentially at large distances, $G_6(y)$ still remain finite.

\begin{figure}[htbp]
\epsfxsize=6cm
\epsfysize=5cm
\centerline{\epsfbox{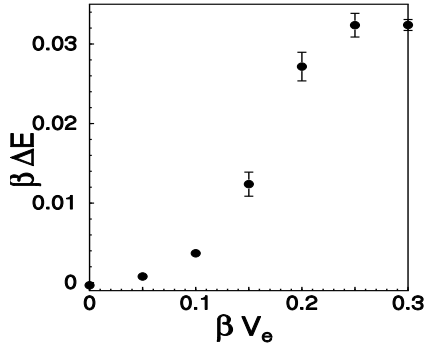}}
\caption{Energy difference between anti-periodic and periodic boundary
conditions}
\label{fig:aperiodic}
\end{figure}

In figure \ref{fig:aperiodic} we have shown the difference in energy
in units of $k_B T$ per particle for anti-periodic and periodic 
boundary conditions. We have considered a system with $L = 10$ in order
to have a
measurable enegy difference because of the large shear.
The system freezes around $\beta V_e = 0.2$. From
figure \ref{fig:aperiodic} we find that there is a non-zero 
elastic energy cost and hence a non-zero shear modulus 
{even before the system freezes to crystalline structure.}
At $\beta V_e = 0.2$, there is a sharp change in $\Delta E$, which
then saturates. 

\section{Conclusion}
We have shown that the {\it modulated liquid} phase 
obtained by inducing density modulation in a colloidal liquid by
subjecting it to  external laser modulations has 
properties intermediate between crystal and liquid; specifically
partial translational order, finite bond-orientational order, and
non-zero rigidity modulus. 
In two dimensional freezing,
a large part of the loss of entropy upon freezing is due to 
the bond-orientational ordering. 
In LIF, the fact that the freezing transition occurs from a 
 partially bond-orientationally ordered
phase to a crystalline phase, helps one to understand why the 
transition becomes more weakly first order
as the external field strength is increased.

\section*{Acknowledgments}
\label{sec.ack}
The authors thank C. DasGupta, R. Pandit
 and S.S. Ghosh for many useful discussions.
We thank SERC, IISc for
computing resources. CD thanks CSIR, India for
financial support.

\end{document}